\documentclass[sn-nature]{sn-jnl}

\usepackage{graphicx}%
\usepackage{multirow}%
\usepackage{amsmath,amssymb,amsfonts}%
\usepackage{amsthm}%
\usepackage{mathrsfs}%
\usepackage[title]{appendix}%
\usepackage{xcolor}%
\usepackage{textcomp}%
\usepackage{manyfoot}%
\usepackage{booktabs}%
\usepackage{algorithm}%
\usepackage{algorithmicx}%
\usepackage{algpseudocode}%
\usepackage{listings}%
\usepackage[version=4]{mhchem}
\usepackage{siunitx}
\usepackage [utf8]{inputenc}
\usepackage[T1]{fontenc}
\usepackage{lineno}

\begin{document}

\title[Article Title]{Superefficient microcombs at the wafer level}

\author[1]{\fnm{Marcello} \sur{Girardi}}

\author[1]{\fnm{\'Oskar B.} \sur{Helgason}}

\author[1]{\fnm{Carmen H.} \sur{L\'opez Ortega}}

\author[1,2]{\fnm{Israel} \sur{Rebolledo-Salgado}}

\author*[1]{\fnm{Victor}\sur{Torres-Company}}\email{torresv@chalmers.se}

\affil[1]{\orgdiv{Department of Microtechnology and Nanoscience}, \orgname{Chalmers University of Technology}, \orgaddress{\city{G\"oteborg}, \postcode{SE-41296}, \country{Sweden}}}
\affil[2]{\orgdiv{Measurement Science and Technology}, \orgname{RISE Research Institutes of Sweden}, \orgaddress{\city{Bor\aa s}, \postcode{SE-50115}, \country{Sweden}}}

\abstract{Photonic integrated circuits utilize planar waveguides to process light on a chip, encompassing functions like generation, routing, modulation, and detection. 
Similar to the advancements in the electronics industry, photonics research is steadily transferring an expanding repertoire of functionalities onto integrated platforms  \cite{Sun2015Single-chipLight}. 
The combination of best-in-class materials at the wafer-level increases versatility and performance  \cite{Snigirev2023UltrafastPhotonics}, suitable for large-scale markets, such as datacentre interconnects  \cite{Jrgensen2022Petabit-per-secondSource, Rizzo2023MassivelyLink}, lidar for autonomous driving  \cite{Rogers2021APlatform}  or consumer health. 
These applications require mature integration platforms to sustain the production of millions of devices per year and provide efficient solutions in terms of power consumption and wavelength multiplicity for scalability.
Chip-scale frequency combs \cite{Chang2022IntegratedTechnologies} offer massive wavelength parallelization, holding a transformative potential in photonic system integration \cite{Shu2022Microcomb-drivenSystems}, but efficient solutions have only been reported at the die level \cite{Xiaoxiao2017MicroresonatorEfficiency, Cuyvers2021LowLaser, Hu2022High-efficiencyGenerators}.
Here, we demonstrate a silicon nitride technology on a 100mm wafer that aids the performance requirements of soliton microcombs in terms of yield, spectral stability, and power efficiency. 
Soliton microcombs are reported with an average conversion efficiency exceeding 50\%, featuring 100 lines at 100 GHz repetition rate. 
We further illustrate the enabling possibilities of the space multiplicity, i.e., the large wafer-level redundancy, for establishing new sensing applications, and show tri-comb interferometry for broadband phase-sensitive spectroscopy. 
Combined with heterogeneous integration of lasers  \cite{Xiang2021LaserSilicon, Xiang20233DPhotonics}, we envision a proliferation of high-performance photonic systems for applications in future navigation systems, data centre interconnects, and ranging.}

\maketitle
Photonic integrated circuits (PICs) are devices that process light on a chip by guiding it through waveguides tailored to perform specific operations \cite{Siew2021}. 
Following the electronic industry footprints, photonic research is transferring an increasing number of functionalities to integrated platforms with the aim of implementing full photonic systems on a chip  \cite{Xiang2022, Bogaerts2020}. 
This is beneficial for the steady deployment of photonic integrated circuits in large-scale markets, such as transceivers for data centre interconnects \cite{Ferraro2023ImecRoadmap} and lidars  \cite{Martin2018PhotonicLiDAR, Chen2023BreakingChaos} for automotive applications. 
These markets require mature integration platforms that can potentially supply millions of devices per year and provide efficient solutions in terms of power consumption, wavelength multiplicity, and optical performance \cite{Xiang2022}.

Silicon is one of the most successful material platforms for photonic integration, with silicon-on-insulator as the standard wafer available in sizes up to 300 mm.
In the last two decades, this platform evolved from research to commercialization \cite{Rahim2019}, and it has been deployed in communication transceivers modules since 2007 \cite{Gunn2007ATransceiver}. 
Silicon photonics developed widely thanks to its similarities with standard complementary metal-oxide semiconductor (CMOS) manufacturing, which kept the cost of PIC fabrication low and guaranteed the volume necessary for scalability in a consumer market. 
However, other platforms such as silicon nitride \cite{Bauters2011Ultra-low-lossWaveguides, Liu2021a}, lithium niobate \cite{Zhu2021IntegratedNiobate, Snigirev2023UltrafastPhotonics} and III-V materials \cite{Arafin2018AdvancedSensing} are being developed, and some are emerging at the commercial level \cite{Munoz2019FoundryMid-Infrared, Ye2023FoundryCircuits} with prototype lines. 
These platforms benefit from the same economy of scale that established SOI as the de facto standard to build the next generation of PICs \cite{Xiang2021a, Roelkens2023Micro-TransferCircuits}.

One of the recent milestones in PICs is the integration of narrow linewidth lasers \cite{Xiang20233DPhotonics} on a \ce{Si3N4} platform that displays ultra-low loss in the telecom frequency range. 
This achievement enables the development of photonic systems on a chip for applications such as spectroscopy \cite{Dutt2018On-chipSpectroscopy}, 
or coherent optical communications \cite{Marin-Palomo2017Microresonator-basedCommunications}. 
However, these applications require a massive wavelength parallelization, which translates into the integration of tens to hundreds of lasers on-chip. 
Such large-scale multiplicity can be easily achieved with integrated microresonator frequency combs (microcombs) \cite{Kippenberg2018DissipativeMicroresonators, Sun2023ApplicationsMicrocombs}.
\begin{figure}[t]%
\centering
\includegraphics[width=\textwidth]{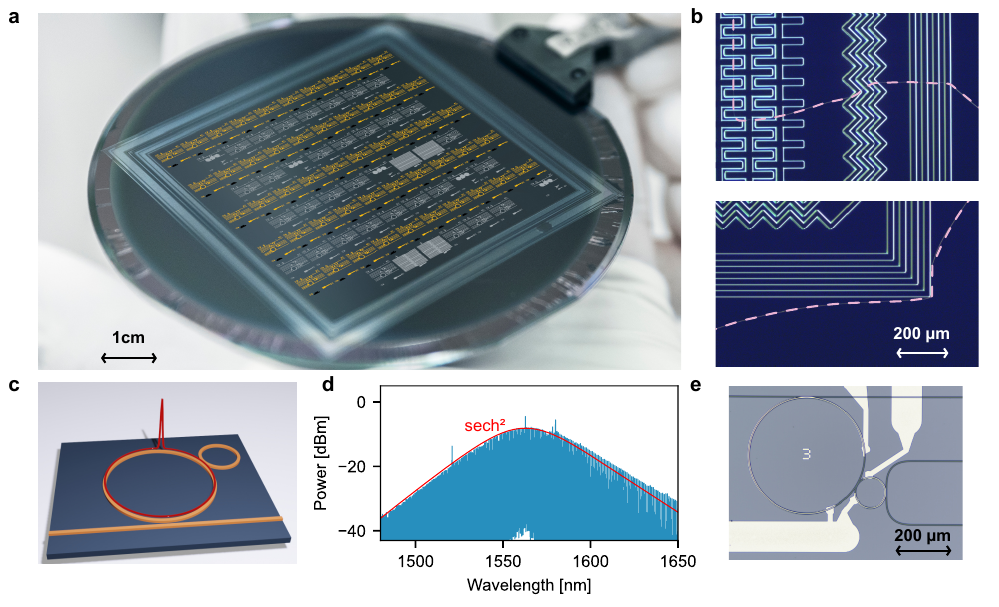}
\caption{{\bf Wafer-scale fabrication of soliton microcombs.} 
\textbf{a.} Picture of a wafer before the singulation process with crack barriers patterned at the edge of the chip area. The 50 chips measured in this work are highlighted in orange. 40 more chips that are not part of this work are present on the wafer (highlighted in white). 
\textbf{b.} The design of the barrier deflects the cracks (highlighted by the dashed violet line) avoiding propagation of the defects in the patterned area, allowing for the wafer-scale processing of ultra-low-loss silicon nitride PICs with high yield in strong optical confinement waveguide geometries.  
\textbf{c.} The soliton comb is generated in the main cavity with the auxiliary cavity introducing a constant phase shift of the pump mode. The shift enables an increased amount of pump power coupled in the main cavity, i.e., a higher power conversion efficiency.
\textbf{d.} One of the power-efficient frequency comb generated with the \(sech^2\) envelop fitting. 
\textbf{e.} Microscope picture of one of the devices with the main and auxiliary ring resonators and the metal heaters used to tune the cavities. 
}\label{fig:vision}
\end{figure}
Dissipative Kerr soliton microcombs rely on a \(\chi^{(3)}\) nonlinear microcavity driven by a continuous-wave laser. 
Soliton pulses come out of the cavity at a rate commensurate to the free spectral range. 
The periodic sequence of the pulse train results, in the frequency domain, into a set of evenly spaced narrow lines underneath a smooth hyperbolic secant shape. 
The spectrum is so broad that hundreds of lines at ultrahigh ($>$100 GHz) repetition rates may be generated. 
Crucially, this is broader than other chip-scale comb technologies relying on semiconductor materials \cite{Hermans2022OnChipSources}, and the repetition rate is higher than state-of-the-art low-noise RF oscillators needed in e.g. electro-optic comb generators \cite{Carlson2018UltrafastControl, Hu2022High-efficiencyGenerators}. 
Soliton microcombs are generated all-optically, and the DC current used to control the cavities has in practice a negligible influence on the overall wall-plug efficiency. 
The efficiency of the comb generation process is crucial to maximize the optical power available on the PIC. 
Recently, efficient solutions have been reported in the literature \cite{Xue2017MicroresonatorEfficiency, Li2022EfficiencyMicrocombs, Boggio2022EfficientBack-coupling, Helgason2022Power-efficientMicrocombs} but they have only been demonstrated at the die level. 
To exploit the full potential of chip-scale comb generators for practical, large-volume applications, it is paramount to address their scalability. 

While silicon nitride has been manufactured with ultralow-loss at the wafer-level before \cite{Liu2021a, Jin2021Hertz-linewidthMicroresonators, Ye2023FoundryCircuits}, this is insufficient to guarantee a reproducible, high-level performance for microcombs. Only reproducible designs that can tolerate the process variation in a standard CMOS line will be able to satisfy a market that requires millions of devices.  
Our fabrication relies on a subtractive manufacturing process where the \ce{Si3N4} is deposited on an oxidized Si wafer with 3 \unit{\um} of \ce{SiO2}. 
A large area is available on the wafer without any pre-patterned structures thanks to the crack barriers positioned on the perimeter (Fig. \ref{fig:vision}a and b). 
This provides high flexibility in the chip dimensions and the design of the devices combined with high fabrication yield.

In this work, we demonstrate that photonic molecule super-efficient microcombs \cite{Helgason2022Power-efficientMicrocombs} can be manufactured on a wafer scale with a fabrication yield of 98\%.
All the tested frequency combs display high power-conversion efficiency - the 25th percentile and the average are 50.3\% and 50.7\%. 
This is to the best of our knowledge the first statistical analysis of chip-scale frequency comb generators of any kind.
These results not only represent a crucial step towards the large-scale deployment of chip-scale frequency combs in mass-market applications but also open new scientific opportunities. 
To display the enabling possibilities of the high reproducibility and yield, we realize tri-comb spectroscopy. Specifically, spectrally phase-sensitive signals over 40 nm bandwidth can be obtained in a highly parallel manner using a common local oscillator comb.  
Overall, these results represent a crucial step towards deploying microcomb technology in large-scale applications, where the CMOS processes can be exploited to fabricate millions of devices.

\bigskip
\noindent
\textbf{Wafer-level fabrication of microcombs} \\
The generation of dissipative Kerr solitons requires a material that displays a balance between nonlinearities, anomalous group velocity dispersion (\(\beta_2\)), and low optical losses \cite{Kippenberg2018DissipativeMicroresonators}. 
\ce{Si3N4} is an excellent platform that has been widely investigated to generate frequency combs in the near-infrared \cite{Chang2022IntegratedTechnologies}. 
The generation of frequency combs on-chip requires careful dispersion engineering of the optical waveguide and in this study, we designed combs operating in the anomalous dispersion regime (\(\beta_2<0\)). 
The stoichiometric \ce{Si3N4} used in our fabrication process displays anomalous dispersion in the telecom C band (1550 nm) when the waveguide core has a thickness larger than 650 nm. 
However, the fabrication of \ce{Si3N4} films above 400 nm presents challenges in terms of crack management because of the high stress accumulated in the film during deposition \cite{Luke2013OvercomingResonators,Xiang2022, Grootes2022CrackDicing}. 
Stress-generated cracks hinder fabrication with high yield, hence stress release patterns are necessary to stop the crack diffusion in the interested area of the wafer. 
These crack barriers consist in structures etched in the bottom cladding and in the Si substrate to dissipate the energy of the crack \cite{Nam2012}. 
They can be as simple as scratches performed with a diamond tip \cite{Luke2013OvercomingResonators}, complex patterns surrounding the photonic devices \cite{Pfeiffer2018PhotonicWaveguides,Wu2020}, or diced at the edges of the interested area \cite{Grootes2022CrackDicing}. 
On our wafers, we etch the crack barrier at the perimeter of the patternable area of the wafer. 
This allows us to redesign the photonic layer after the material is deposited. Our crack barriers are etched in the bottom cladding and provide a crack-free area of 56x56 $\text{mm}^2$ on a 100 mm wafer (\ref{fig:vision}a). 
The cracks are efficiently stopped by the crack barriers, as shown in Fig. 1b. 
We can easily accommodate 100 chips with size 5x5 \(\text{mm}^2\)  in the crack-free area. 

We fabricated ninety chips on the wafer, fifty of which are dedicated to studying the yield of power-efficient frequency combs. 
The chips are organized in five columns and ten rows, to cover the entire crack-free area, as displayed by the orange pattern highlighted in Fig. \ref{fig:vision}a. 
Each chip included one device, where two ring resonators are coupled to generate a power-efficient comb, similar to \cite{Helgason2022Power-efficientMicrocombs}. 
A microscope picture of a representative device is reported in Fig. \ref{fig:vision}e. The device consists of a main cavity, with a radius of 227.82 \(\mu m\), and an auxiliary cavity with a radius of 60 \unit{\um}. 
The waveguide width is 1.8 \(\mu m\) for the bus coupler, the main cavity, and the auxiliary cavity. 
The two resonators are designed to have a free spectral range close to 100 GHz and 380 GHz respectively. 
The input and output of the chip have spot-size converters designed to couple to a lensed fiber with a mode field diameter of 2.5 \(\mu m\). 
In Fig. \ref{fig:vision}d we report one of the combs generated with the expected \(\text{sech}^2\) fitted curve. The strong attenuation of the pump is due to the high conversion efficiency; it showcases the potential of this type of frequency comb, where the pump does not need to be filtered out.

\begin{figure}[t!]%
\centering
\includegraphics[width=\textwidth]{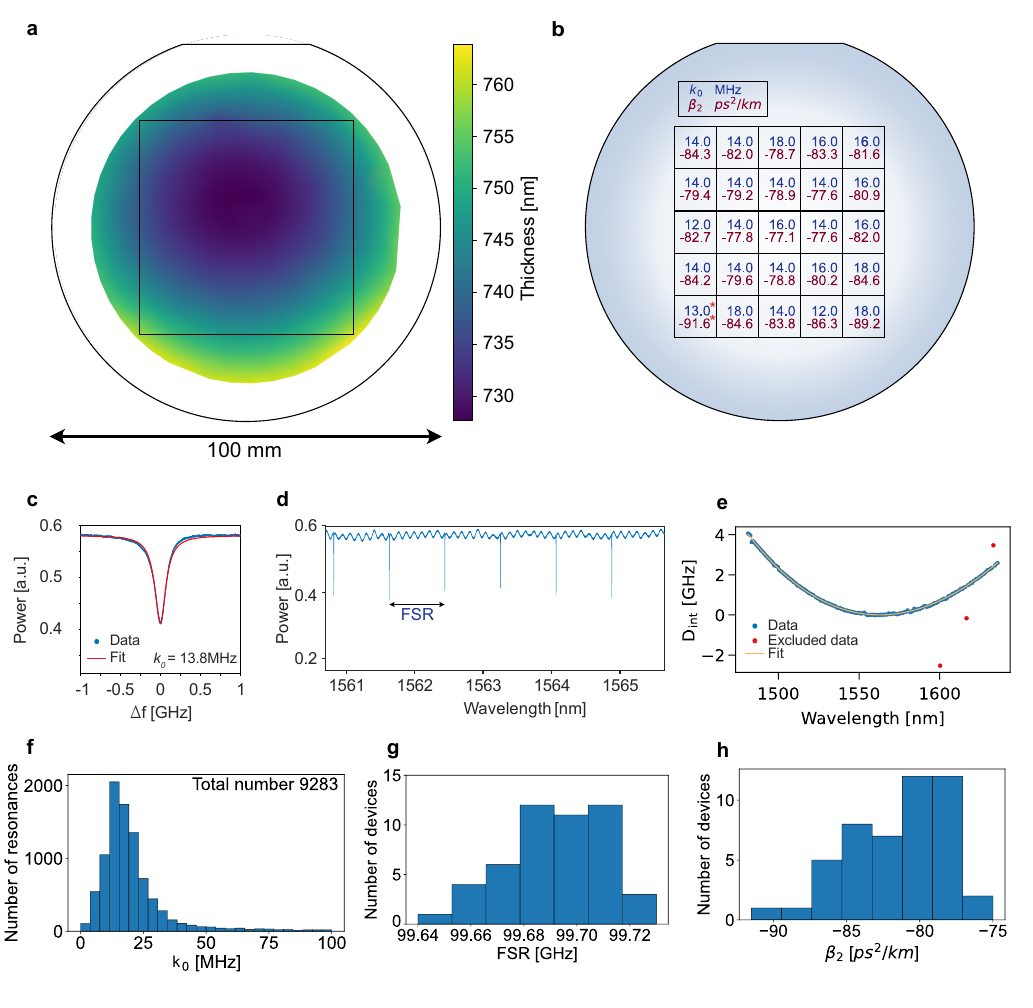}
\caption{{\bf Linear characterization of the coupled ring resonators.}
\textbf{a.} Thickness map of the \ce{Si3N4} layer deposited via low-pressure chemical vapour deposition with the outline of the wafer (circle) and the area dedicated to chip fabrication (square). 
\textbf{b.} Uniformity analysis on the 100 mm wafer. The most probable value of the intrinsic linewidth \(k_0\) and measured group velocity dispersion \(\beta_2\) are reported as the mean between two adjacent chips. The red asterisk marks the cell with the faulty chip. 
\textbf{c.} Representative resonance of the main cavity and the fitted resonance shape
\textbf{d.} Portion of the transmission spectrum of one of the devices measured displaying the FSR.
\textbf{e.} Typical plot of the integrated dispersion, with highlighted in red the points excluded due to the interaction with the auxiliary cavity.
\textbf{f.} Histogram of the intrinsic linewidth \(k_0\) of all the resonances measured on the wafers. 
\textbf{g, h.} Histograms of the free spectral range and the group velocity dispersion of the main cavity.}\label{fig:linear}
\end{figure}

The deposition of \ce{Si3N4} via low-pressure chemical vapour deposition yields an inhomogeneous thickness on the wafer \cite{Gumpher2003LPCVDControl}. 
On our wafer, the thickness variation is 2.4\% across the wafer and 1.4\% within the crack barriers. 
This variation affects the core geometry of the waveguides fabricated in different areas of the chip, hence it modifies the dispersion. 
This can have a great impact on the performance of the frequency comb, as we will discuss in the frequency comb generation section. 
In a research environment, the design of every chip can be tailored to the local thickness of the material, but this might be not acceptable in large-scale manufacturing. 
To demonstrate that the power-efficient soliton generation is tolerant to variation in the fabrication parameters, we replicate the same design on every chip.

\bigskip
\noindent
\textbf{Linear characterization} \\
To assess the fabrication yield we started with the linear characterization of the devices. 
We assessed the optical loss, the free spectral range (FSR), and group velocity dispersion (\(\beta_2\)) via swept wavelength interferometry \cite{Twayana2021} (see details in Methods). In Fig. \ref{fig:linear}f, we report the histogram of the intrinsic linewidth (\(k_0\)) for the processed resonances of all the devices measured, corresponding to over 9000 resonances. 
The most probable value is 13.5 MHz which corresponds to a quality factor of 14.2\(\times10^6\)  and an equivalent propagation loss of 2.6 dB/m. 
The fabrication yield is 98\%, with a single chip failure due to residual particles on the bus waveguide. 
The chip in question is on the bottom left side of the pattern, as illustrated by the red marker (\textcolor{red}{*}) in Fig. \ref{fig:linear}b, where we report the distribution of the linewidth and the \(\beta_2\) across the wafer. 
Each value is the average of two adjacent chips on the same column. 
As expected, the \(\beta_2\) value changes across the wafer and it is higher in the centre, where the \ce{Si3N4} thickness is lower. 
This is also illustrated in the map in Extended Data Fig. 2c, where we report the \(\beta_2\) value of every device. The histograms in Fig. \ref{fig:linear}g and \ref{fig:linear}f show the distributions of the FSR and \(\beta_2\) values respectively. 
With a mode solver simulation, we calculated the expected parameters for this waveguide geometry and we obtained that the most probable value of the FSR and \(\beta_2\) are respectively 99.7 GHz and -72 \(\text{ps}^2\)/km, in line with the experimental results. 
This analysis illustrates the ability to perform accurate designs of ultra-high-Q resonators with our silicon nitride process.
\begin{figure}%
\centering
\includegraphics[width=\textwidth]{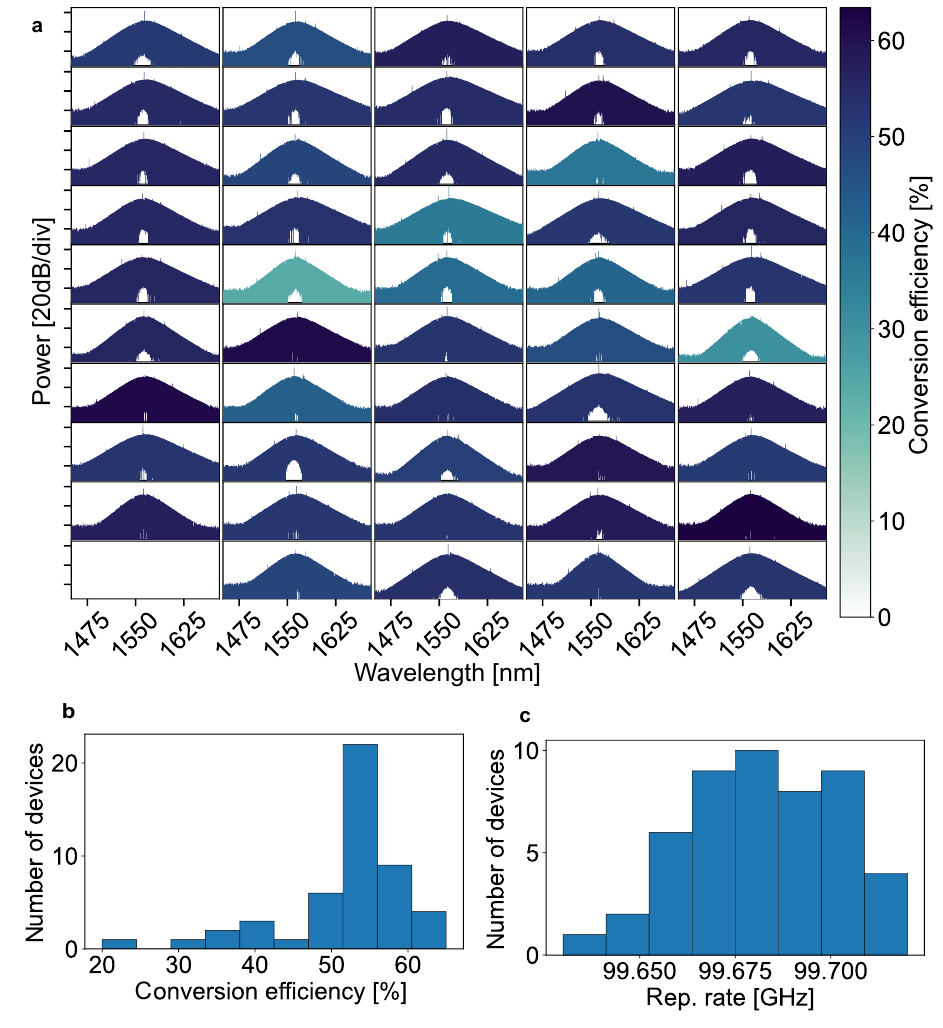}
\caption{{\bf High yield generation of power-efficient microcombs.}
\textbf{a.} Spectra of every soliton comb generated across the wafer. The colour of the plot represents the power conversion efficiency of the soliton. 
\textbf{b, c.} Histograms of the power conversion efficiency and the repetition rate of the generated soliton comb.}\label{fig:microcomb map}
\end{figure}

\bigskip
\noindent
\textbf{Frequency comb generation} \\
After the linear characterization of the devices, we generated a frequency comb in each device as described in the Methods. 
In Fig. \ref{fig:microcomb map}a we report the 49 combs generated with power conversion efficiency ranging between 24 and 63\%. 
The colour of the comb represents the conversion efficiency, as displayed in the colour bar on the right of the plot. 
The power conversion efficiency is on average 50.7\% and the 25th percentile is 50.3\%. 
Indeed, the majority of the devices display a conversion efficiency greater than 50\% with the most probable value at 53.7\%. 
To the best of our knowledge, the maximum conversion efficiency of 63\% obtained in these measurements is the highest conversion efficiency recorded in a single DKS driven by a continuous-wave laser.

The power conversion efficiency is influenced by multiple factors. 
We investigated the correlation between the power conversion efficiency with \(k_0\), \(\beta_2\), and coupling between cavities (\(k_{aux}\)) in the plots in Extended Data Fig. \ref{fig:microcomb map extended}a-c. 
We see a correlation between the group velocity dispersion and the power conversion efficiency, partially confirmed by the simulation reported in Fig. \ref{fig:microcomb map}d-f. 
The coupling with the auxiliary cavity does not display a significant correlation with the comb efficiency, while the intrinsic linewidth shows that the least efficient combs also display higher intrinsic loss. 
Overall, the simulation analysis predicts that the photonic molecule microcomb is resilient to variations in design parameters. 
Note however, that the simulation does not capture all physical features of the resonator, such as wavelength-dependent linewidth, which may be the reason for some of the measured combs exhibiting less than 40\% efficiency.

As expected from the FSR measurement, the repetition rate changes across the wafer with a standard deviation of 20 MHz (see Extended Data Fig. \ref{fig:microcomb map extended}a).
This suggests that we can design the repetition rate with high precision, a crucial aspect for dual-comb spectroscopy \cite{Duran2015UltrafastInterferometry}, comb synchronization \cite{Kim2021SynchronizationCombs}, and advanced coherent communication systems \cite{Geng2022CoherentMicrocombs} relying on phase-coherent transmitter and receiver combs with repetition rate differences within a few MHz.

\bigskip
\noindent
\textbf{Phase-sensitive tri-comb interferometry} \\
The high yield of power-efficient solitons presents unique opportunities for chip-integrated applications such as spectroscopy \cite{Dutt2018On-chipSpectroscopy}  and range measurement \cite{Trocha2018UltrafastCombs, Suh2018SolitonMeasurement, Riemensberger2020MassivelyMicrocomb}.
Moreover, the generation of frequency combs through a photonic molecule configuration provides deterministic generation, stable operation, and a relatively flat spectral distribution. 
As proof of concept, we demonstrate how these capabilities can be used together. 
\begin{figure}%
\centering
\includegraphics[width=\textwidth]{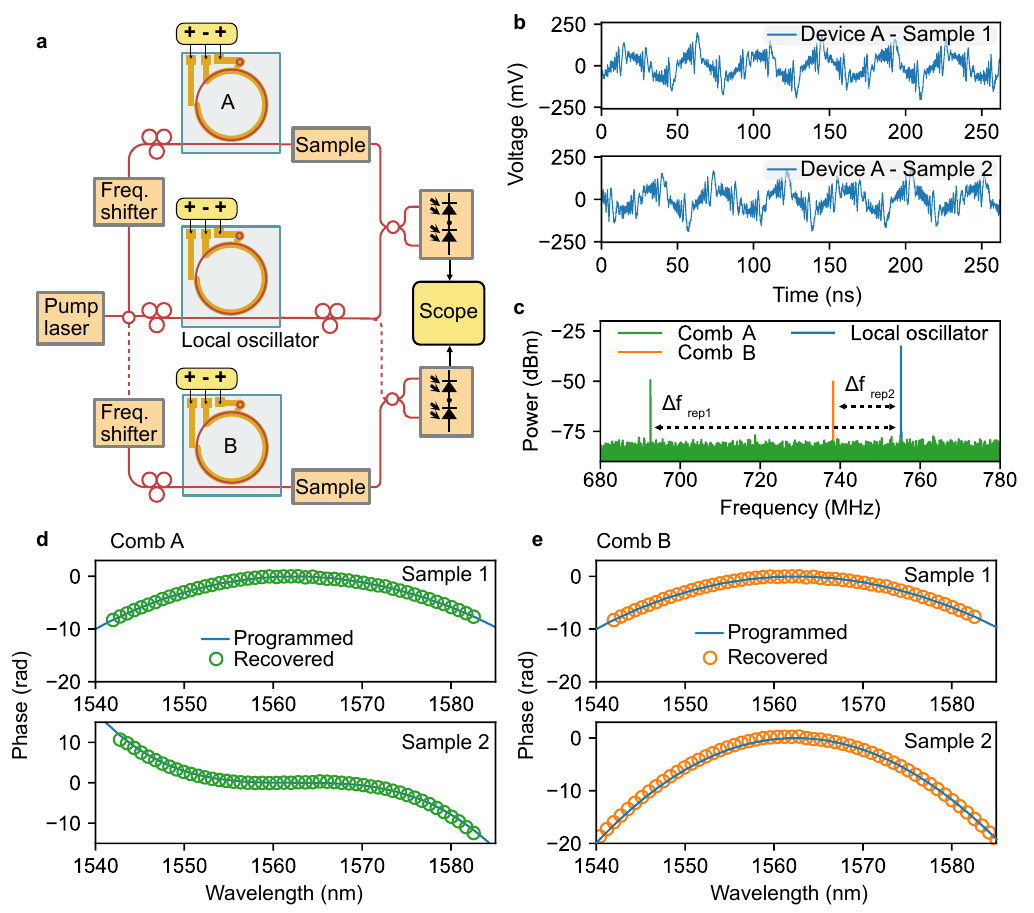}
\caption{{\bf Tri-comb interferometry.}
\textbf{a.} schematic of the experimental setup for the microcomb interferometry. 
\textbf{b.} Interferogram traces of the beating between comb A and the local oscillator acquired with the of comb A with the two samples. 
\textbf{c.} Downconverted beat notes of the local oscillator (blue), comb A (green), and comb B (orange). 
\textbf{d, e.} Plots of the programmed phase (solid line) and the recovered phase (empty dots).}\label{fig:tricomb spectr}
\end{figure}

In our demonstration, we propose a multi-comb interferometry scheme that enables parallel sampling of multiple devices under test. 
Based on multi-heterodyne detection, we employed a single local oscillator for the characterization of amplitude and phase in a sequential manner of two samples. 
The experimental setup is illustrated in Fig. \ref{fig:tricomb spectr}a. 
We utilize three power-efficient combs: one for implementing the optical sampling (local oscillator) and the other two (comb A and comb B) are employed for consecutive measurements with the samples. 
The generation of coherent photonic molecule microcombs follows the same approach described in the Methods. 
To establish mutual coherence, we employ the same pump laser at a wavelength of 1562.3 nm. The two combs were coupled to an optical pulse shaper \cite{Leaird2009Dual-combCharacterization,Duran2015UltrafastInterferometry} to simulate multiple different samples by adding a phase profile to the transmitted signal. 
The combs were combined with the local oscillator to obtain the interferograms reported in Fig. \ref{fig:tricomb spectr}b. 
The period corresponds to the difference in repetition rate between the local oscillator and the probe combs, which are \(\sim\)15 ns (\(1/\Delta f_{rep1}\)) and \(\sim\) 58 ns (\(1/\Delta f_{rep2}\)) for comb A and B, respectively.  
The electro-optic down-converted repetition rate of the solitons is shown in Fig. \ref{fig:tricomb spectr}c, where \(\Delta f_{rep1}\) is equal to 61.35 MHz and \(\Delta f_{rep2}\) is equal to 17.23 MHz. 
Importantly, the laser pump generating the combs is not filtered out before detection. 
This is aided by the high conversion efficiency, which drastically reduces the intensity of the pump in the transmitted spectrum, hence allowing a simpler detection scheme.

The time-domain signal is processed using Fourier analysis to retrieve the spectral information, where an RF comb spaced by \(\Delta f_{rep1,2}\) is obtained. 
The post-processing of the signal was implemented using coherent averaging and monitoring the drift of \(\Delta f_{rep1,2}\) \cite{Trocha2018UltrafastCombs}. 
The number of beat notes is limited by the bandwidth of the optical filter programmed in the pulse shaper, which covers 45 nm. 
Fig. \ref{fig:tricomb spectr}d shows the spectral phase information retrieved using the two solitons in consecutive measurements. 
These results were obtained when the pulse shaper was programmed for quadratic and cubic phase profiles. 
The first phase filter is the same for both combs A and B, and a different second filter is encoded in each comb. 
The empty circles represent the retrieved spectral phase at the peaks of the RF combs and the lines are the functions programmed. 
The standard deviation of the residuals between the retrieved and the encoded phase is 0.07 rad for the first filter and 0.25 rad for the second filter,
illustrating that the technique can recover with high spectral sensitivity multiple probe samples derived from a single continuous-wave laser. 

\bigskip
\noindent
\textbf{Conclusion and outlook}\\
In conclusion, we have reported the first statistical analysis of wafer-level manufacturing of chip-scale frequency comb generators. 
We have demonstrated 98\% yield with microcombs displaying a power conversion efficiency $>$50\%. 
These results are enabled by advances in ultra-low-loss silicon nitride processing and photonic-molecule engineering. 
The manufacturing process is subtractive and compatible with large-scale foundries \cite{Rahim2019}. 
When combined with heterogeneous integration of pump laser modules \cite{Xiang20233DPhotonics}, these results pave the way for mass-market applications of chip-scale microcomb modules, such as datacentre interconnects in co-package optics or autonomous driving. 
Furthermore, the ability to reproduce with high fidelity nearly identical copies of high-performance microcombs might enable new modalities of sensing and spectroscopy that benefit from the economics of scale provided by silicon photonics. 
\backmatter

\section*{Methods}
\textbf{Device fabrication.}
The process starts with the thermal oxidation of the silicon wafers to obtain a 3 \unit{\um} bottom cladding. 
The crack barriers are patterned with UV mask lithography and etched in the thermal oxide via buffered oxide etching. 
The stoichiometric \ce{Si3N4} is deposited via low-pressure chemical vapour deposition (LPCVD) in two steps with thermal cycling at 1100\unit{\degreeCelsius} to attain a crack-free layer. 
The devices are patterned via electron beam lithography (EBL) and etched with \ce{CHF3}-based inductive coupled plasma reactive ion etching (ICP-RIE). 
Afterwards, an annealing step in Ar at 1200\unit{\degreeCelsius} is implemented to remove residual N-H bonds. 
The top cladding is deposited via LPCVD with TEOS precursor to obtain 3 \unit{\um} of \ce{SiO2} that isolates the waveguides from the metal heaters. We fabricate the thermal phase shifter by lift-off of 200 nm of TiPt evaporated on the substrate. 
The pattern is defined via maskless lithography. 
The optical facets are defined by \ce{SiO2} ICP-RIE and the chips are singulated via deep silicon etching.

\bigskip
\noindent
\textbf{Linear characterization of coupled ring resonators and data processing.}\\
The linear characterization of the devices is performed via swept wavelength interferometry. 
A laser is coupled to the chip and the wavelength is tuned between 1480 nm and 1640 nm. 
The scanning wavelength is calibrated with a self-referenced frequency comb to eliminate the uncertainty given by the nonlinear sweeping of the laser \cite{DelHaye2009FrequencyDispersion}. 
From the response of the resonator, we extract the resonance frequency location and the resonance linewidth. 
We fit each resonance with a Lorentzian curve and with a split resonance fitting that includes the coupling to the backward propagating mode \cite{Pfeiffer2018PhotonicWaveguides}. 
From both fittings, we can extract the intrinsic and extrinsic linewidths, respectively \(k_0\) and \(k_{ex}\). 
The split resonance fitting also provides the coupling rate to the backward propagating mode, \(k_c\). 
We picked the best fitting between split resonance and Lorentz by selecting the result with the highest $r^2$. Moreover, if the best $r^2$ is below 0.96, we discarded the fit. 
This method discarded \(\sim\)2\% of all the resonances measured in this study.

We calculated the FSR and group velocity dispersion by fitting a fourth-order polynomial curve to the resonance relative locations $\omega_\mu-\omega_0$, where $f_\mu=\omega_\mu/2\pi$ is the location of the $\mu-th$ resonance, and $f_0=\omega_0/2\pi$ is the frequency corresponding to a wavelength of 1563 nm. 
The relative resonance location can be expressed as:
\begin{equation}
    \omega_\mu-\omega_0=D_1\mu+(D_2 \mu^2)/2+(D_3 \mu^3)/6+(D_4 \mu^4)/24
\end{equation}
where $D_1/2\pi$ is the FSR and $-D_2/D_1^3R$ is the $\beta_2$, with R the radius of the resonator. 
The dispersion profile can be easily visualized with $D_{int} (\mu)=\omega_\mu-\omega_0-D_1\mu$. 
In this profile, it is easy to identify the avoided mode crossings that are normally caused by the coupling of the fundamental mode with higher-order modes. 
In our case, we have an additional contribution to the avoided mode crossing given by the linear coupling between the main and auxiliary cavities in the device. 
These avoided mode crossings are much stronger than the ones with the higher-order modes and have a strong effect on the calculation of the dispersion. 
For this reason, we removed these resonances from the polynomial fitting of $D_{int}$ by applying a filter based on the fitting $r^2$. 
We calculated the fitting and removed the point with maximum residual until the $r^2$ value reached the selected target of $1-5\time10^{-11}$. 
In Data Extended Fig. \ref{fig:linear_extended}e we report a typical plot of $D_{int}$ with the three avoided mode crossings removed by the strategy described.

\bigskip
\noindent
\textbf{Power-efficient frequency comb initiation.}
We start by identifying one resonance between 1561 nm and 1565 nm which displayed an avoided crossing with the resonance of the auxiliary cavity. 
This is done by monitoring a resonance of the main cavity and tuning the heater of the auxiliary cavity until the main and auxiliary cavity resonances display an avoided mode crossing. 
To initiate the single soliton comb, we followed the procedure described in  \cite{Helgason2022Power-efficientMicrocombs}.

To achieve the power-efficient frequency combs reported in Fig. \ref{fig:microcomb map}a, we generated the single soliton comb at \(\sim\)15 dBm on-chip power and then decreased the input to \(\sim\)11.5 dBm. 
At this point, the transmitted pump power is drastically reduced, and the pump is much closer to the $\text{sech}^2$ envelope of the single soliton state. The conversion efficiency can be further improved by operating on the pump detuning.

\bigskip
\noindent
\textbf{Tri-comb interferometry }
To establish mutual coherence among the frequency combs, we employed the same pump laser at a wavelength of 1562.3 nm. 
The pump of combs A and B was frequency-shifted 23 MHz by means of an acousto-optic modulator to avoid ambiguity in the down-converted beat notes. 
As the pump wavelength remained fixed throughout the measurements, the integrated heaters were utilized to precisely align the pump resonances of combs A and B. 
The generated solitons were coupled to an optical pulse shaper that allowed us to impart arbitrary phase profiles to the spectrum to simulate the samples. 
Finally, the solitons carrying the information of the samples were combined with the local oscillator on a 50/50 fiber coupler. 
The interference was measured with a real-time scope using a balanced photodetector with a 5 GHz bandwidth followed by an RF amplifier.

\subsection*{Data availability}
The data used to produce the plots within this paper will be available on Zenodo at the time of publication.

\subsection*{Acknowledgments}
The devices demonstrated in this work were fabricated at Myfab Chalmers.
This work was supported by funding from the European Research Council (projects CoG 771410 and PoC 101064463), 
the Swedish Research Council (projects 2016-06077 and 2020-00453),
the Swedish Foundation for Strategic Research (FID 16-0011), 
and the Knut and Wallenberg Foundation (KAW 2018.0090). 

\subsection*{Authors contributions}
O.B.H. designed the lithography masks. 
M.G. developed the process and fabricated the samples.
M.G., C.H.L.O., O.B.H. performed the linear characterization.
O.B.H. designed the comb initiation. O.B.H., M.G., and C.H.L.O. measured the combs.
I.R.S. performed the tri-comb experiment. 
M.G., I.R.S., O.B.H., and V.T.C. wrote the manuscript with inputs from C.H.L.O. and V.T.C. supervised the project.

\subsection*{Competing interests}
M.G., O.B.H., and V.T-C. are co-founders of Iloomina AB. Others declare no conflict of interest.
\clearpage

\setcounter{figure}{1} 
\renewcommand{\figurename}{Extended Data Fig.}
\begin{figure}%
\centering
\includegraphics[width=\textwidth]{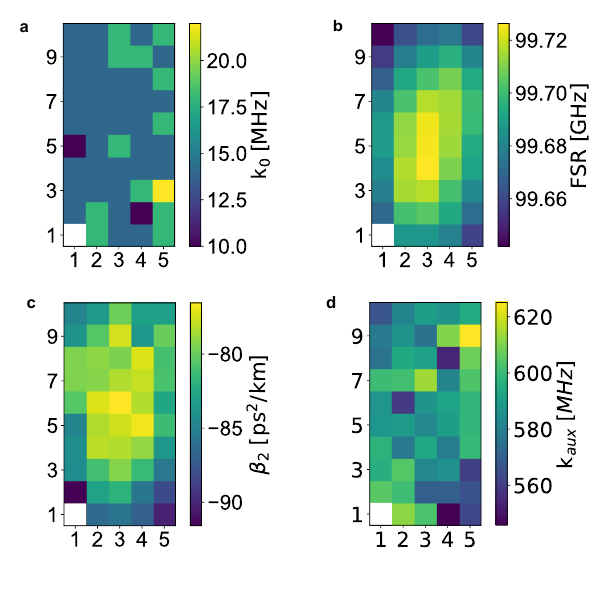}
\caption{
{\bf Parameter distribution across the wafer.}
Maps of the most probable value of the 
intrinsic linewidth (\textbf{a}) 
free spectral range (\textbf{b}) 
group velocity dispersion (\textbf{c}) 
and coupling rate between the main and auxiliary cavities (\textbf{d}) at the pump frequency for the frequency comb generation. 
}\label{fig:linear_extended}
\end{figure}

\begin{figure}
    \centering
    \includegraphics[width=.99\textwidth]{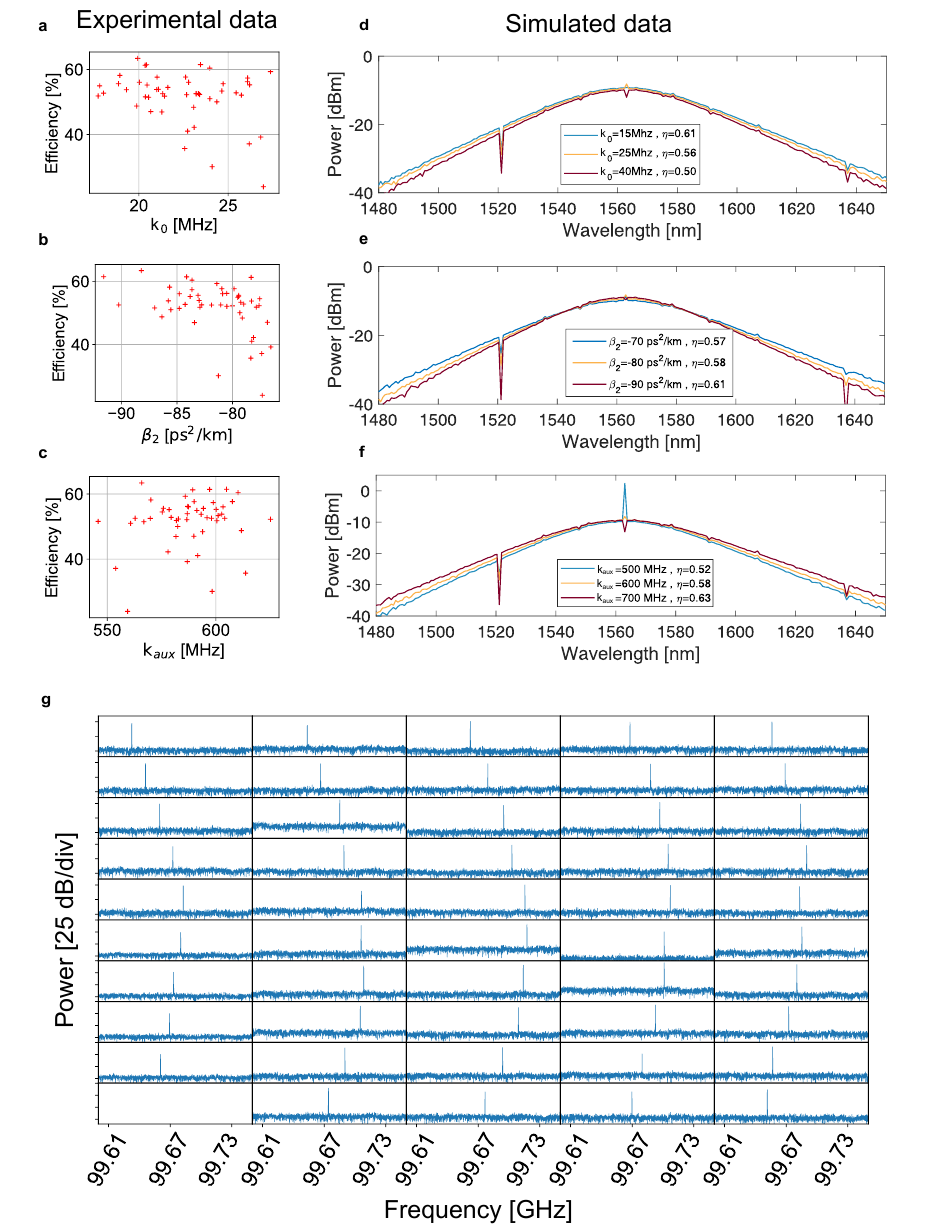}
    \caption{
    \textbf{a, b, }
    and \textbf{c.} Scatter plots of the power conversion efficiency versus the average intrinsic linewidth (\textbf{a}), the group velocity dispersion (\textbf{b}) and the coupling rate between the two cavities (\textbf{c}). 
    \textbf{d, e,} and \textbf{f.} Plots of the simulated combs changing the intrinsic linewidth (\textbf{d}), the group velocity dispersion (\textbf{e}) and the coupling rate between the two cavities (\textbf{f}). The legend of the plot reports the simulated power conversion efficiency. 
    \textbf{g.} Plot of the repetition rate beat note for every comb measured.}
    \label{fig:microcomb map extended}
\end{figure}

\end{document}